\documentclass[%
 reprint,
superscriptaddress,
nofootinbib,nobibnotes,
amsmath,amssymb,aps, prl 
]{revtex4-1}

\usepackage[normalem]{ulem}
\usepackage{siunitx}
\usepackage{graphicx}
\usepackage{dcolumn}
\usepackage{bm}
\usepackage{epstopdf}
\usepackage{bbm}
\usepackage{amsmath}
\usepackage{color}
\usepackage[hidelinks]{hyperref}

\makeatletter
\newcommand{\mypm}{\mathbin{\mathpalette\@mypm\relax}}
\newcommand{\@mypm}[2]{\ooalign{%
		\raisebox{.1\height}{$#1+$}\cr
		\smash{\raisebox{-.6\height}{$#1-$}}\cr}}
\makeatother

\makeatletter
\newcommand*{\citenumns}[2][]{%
	\begingroup
	\let\NAT@mbox=\mbox
	\let\@cite\NAT@citenum
	\let\NAT@space\NAT@spacechar
	\let\NAT@super@kern\relax
	\renewcommand\NAT@open{}%
	\renewcommand\NAT@close{}%
	\cite[#1]{#2}%
	\endgroup
}
\makeatother

\usepackage{commath}
\usepackage{tabularx}

\usepackage{booktabs}

\begin{document}

\preprint{APS/123-QED}

\title{Dirac imprints  on the $g$-factor anisotropy in graphene}

\author{M. Prada}%
\thanks{These authors contributed equally.}
\affiliation{%
	HARBOR, Universit\"at Hamburg, Luruper Chaussee 149,  22761 Hamburg, Germany\\
}%
\author{L. Tiemann}
\thanks{These authors contributed equally.}
\author{J. Sichau}
\thanks{These authors contributed equally.}
\affiliation{%
CHyN, Universit\"at  Hamburg, Luruper Chaussee 149, 22761 Hamburg, Germany\\
}%
\author{R.H. Blick}%
\affiliation{%
CHyN, Universit\"at  Hamburg, Luruper Chaussee 149, 22761 Hamburg, Germany\\
}%
\affiliation{
Materials Science and Engineering, University of Wisconsin-Madison, 1509 Univ. Ave., Madison, WI--53706, USA\\}

\date{\today}
\begin{abstract}
Dirac electrons in graphene are to lowest order spin 1/2 particles, owing to the orbital symmetries at the Fermi level. 
However, anisotropic corrections in the $g$-factor appear due to the intricate spin-valley-orbit coupling of chiral electrons. 
We resolve experimentally the $g$-factor along the three orthogonal directions in a large-scale graphene sample. 
We employ a  Hall bar structure with an external magnetic field of arbitrary direction,  
and extract the effective $g$-tensor via resistively-detected electron spin resonance. 
We employ a theoretical perturbative approach to  identify the intrinsic and extrinsic spin orbit coupling and obtain 
a fundamental parameter inherent to the atomic structure of $^{12}$C, commonly used in ab-initio models.
\end{abstract}


\maketitle
One of the great triumphs of the Dirac relativistic theory for the electron was the prediction of the $g$-factor with the value 
$g_0\simeq 2$ \cite{dirac1928}. 
As a major departure from previous quantum theories, Dirac's equation describes indeed spin 1/2  particles with 4-component spinors or bispinors,  allowing to introduce the concepts of chirality and helicity. Chirality is an inherent property of the particle,  whereas helicity depends on its momentum: namely, it is positive (negative) when the momentum aligns (anti)parallel to the spin. In the massless limit,  both qualities are related \cite{greiner}: positive chirality corresponds to positive helicity and vice versa. 

The linear dispersion at the Fermi level of graphene is cognate  with the Dirac cones of the massless relativistic particles \cite{novoselov2005two}, 
motivating extensive research  towards the parallelism with relativistic quantum mechanics in a solid-state material \cite{katsnelson2006chiral,geim2010rise,Nair08,regan,golub,GIULIANI2012461}.
The inherent chirality  of the  Dirac carriers leads to  a topologically non-trivial band structure \cite{kane2005quantum,kane2005z,jonas}.  
Although the carriers are, to lowest order, spin 1/2 particles, their chirality 
induces a coupling of spin, valley and orbital degrees of freedom \cite{prada}. 
{Here, we address this particular coupling that appears as a measurable $g$-factor anisotropy.}

From a theoretical perspective, the Zeeman Hamiltonian that describes the interaction with an applied field is given by the sum of the contributions of the 
orbital and spin angular momentum, $L$ and $S$, respectively \cite{weil}, 
\[
\hat H_{\rm Z} = \mu_B\vec B(\hat{\vec L} + g_0\hat{\vec S}), 
\]
{with $g_0$ representing the pure $g$-factor, $\mu_B$ the Bohr magneton and $\vec B$ an external magnetic field.} 
On the other hand, the  \emph{effective} spin model,  commonly employed experimentally to describe the Zeeman energy, 
includes an {\it effective} $g$  tensor and  {\it  fictitious} spin operators \cite{bloch,mostafa,rudo1},
\[
\hat H_{\rm eff} = \mu_B\vec B  \tilde{\bm g} \hat{\vec S}. 
\]
$\tilde{\bm g}$  must be constructed such that the energies obtained with the effective spin Hamiltonian
capture  the corrections due to the internal molecular orbital angular 
momentum. 
In electron-spin resonance (ESR) experiments, this internal structure modifies the 
strength of an external field necessary to meet the resonant condition \cite{Slichter1990}: 
\begin{equation}
h\nu = \mu_B\langle \vec B (\hat{\vec L} + g_0\hat{\vec S})\rangle  = 
\mu_B\langle \vec B \tilde{\bm g} \hat{\vec S}\rangle, 
\label{eqR}
\end{equation}
where $\tilde{\bm g}$  is a tensor that contains the effective (or experimental) $g$-factors measured with the 
field along the corresponding directions and $\langle.\rangle$ indicates the expectation value. 
Since the $g$-tensor is diagonal along the crystallographic directions, the angular dependence for the 
general rhombic symmetry 
can be expressed in terms of 
 $g_{xx}$, $g_{yy}$ and $g_{zz}$ \cite{mostafa}: 
\begin{equation}
g(\theta,\varphi) =\sqrt{g_{zz}^2\cos^2{\theta} + g_{yy}^2\sin^2{\theta}\sin^2{\varphi} + g_{xx}^2\sin^2{\theta}\cos^2{\varphi} }, 
\label{eqgeff}
\end{equation}
for an arbitrary magnetic field with axial and  azimuthal  angles $\theta$, $\varphi$. 
In this letter we resolve experimentally the effective $g$-factor along the three main directions 
in a mesoscopic graphene sample, $ g_{\alpha\alpha} = g_0 +  \Delta g_{\alpha\alpha}$,  
whereas the corresponding theoretical correction is evaluated perturbatively  via 
the expectation value of the angular momentum, 
$\Delta g_{\alpha\alpha} = \langle \hat 
L_\alpha\rangle$, $\alpha = x,y,z$.  
We employ a microscopic perturbative model to obtain  $\Delta g_{\alpha\alpha}$ in terms 
of atomic parameters \cite{prada}. 
We then compare {these theoretical} values {to} our experimentally obtained {$g$-factors} and extract the 
{atomic spin-orbit coupling (SOC) corrections}. 

It is commonly accepted that near the Dirac points (DPs)  the eigenstates are described by $\pi$-orbitals near the Fermi edge \cite{neto2009electronic}.
The conduction and valence bands, to lowest order, are linear in momentum, with the corresponding {\it chiral} states 
given in terms of the main ($p_z$-orbital) contribution at sublattices $A$ and $B$ \cite{neto2009electronic,katsnelson2012graphene}: 
\begin{equation}
|\varphi^{(0)}_\pm\rangle \simeq 
c_A|p_z^A\rangle 
+ c_B|p_z^B\rangle, \ \frac{c_B}{c_A} = \pm e^{i\varphi_q \tau} , 
\ \varphi_q= \arctan{\frac{q_y}{q_x}}. 
\label{eq0}
\end{equation}
where the sign $\pm$ labels the conduction band (CB) and valence band (VB), respectively. 
These DPs are the celebrated $K$ and $K^\prime$ points, which are assigned the valley index, $\tau$ = 1 and  -1, 
respectively, and  $\vec  q = (q_x,q_y)$ is the small vector off the nearest DP. 
 The chirality-preserving Kane-Mele intrinsic SOC term \cite{kane2005quantum}, 
$\hat H_{\rm KM} = \tau\lambda_{\rm I} \hat s_z \hat \sigma_z,$  with $\hat s_z$, $\hat \sigma_z $ being the 
Pauli matrices representing the electron spin and sublattice-spin, respectively, leads to the spin Hall effect 
and a measurable intrinsic SOC gap $\Delta_{\rm I} = 2\lambda_{\rm I}$ \cite{jonas,singh2020sublattice,BanszerusPRL20}. 
As we will show, the intrinsic SOC leads as well  to {\it chiral} spin-valley orbit coupling and additional corrections to the measured $g$-factor.
To lowest order, the  quasiparticle eigenenergies are given by: 
\begin{equation}
\varepsilon_\pm = \pm\sqrt{ (\hbar v_F q)^2 + \lambda_{\rm I}^2  }. 
\label{eqGS}
\end{equation}
\textcolor{black}{with $v_F$ being the Fermi velocity.}
The axial symmetry of the $p_z$-orbitals involves $\langle \hat L_\alpha\rangle =0$, and hence, 
the $g$-factor at lowest order is that of free electrons. 
Dominant corrections to the $g$-factor are due to 
(i) band hybridization, 
(ii) atomic SOC, 
(iii) Bychkov-Rashba effect and (iv) structural SOC, which we consider in the following. 

\begin{figure}[!hbt]
\includegraphics[angle=0, width=.60\linewidth]{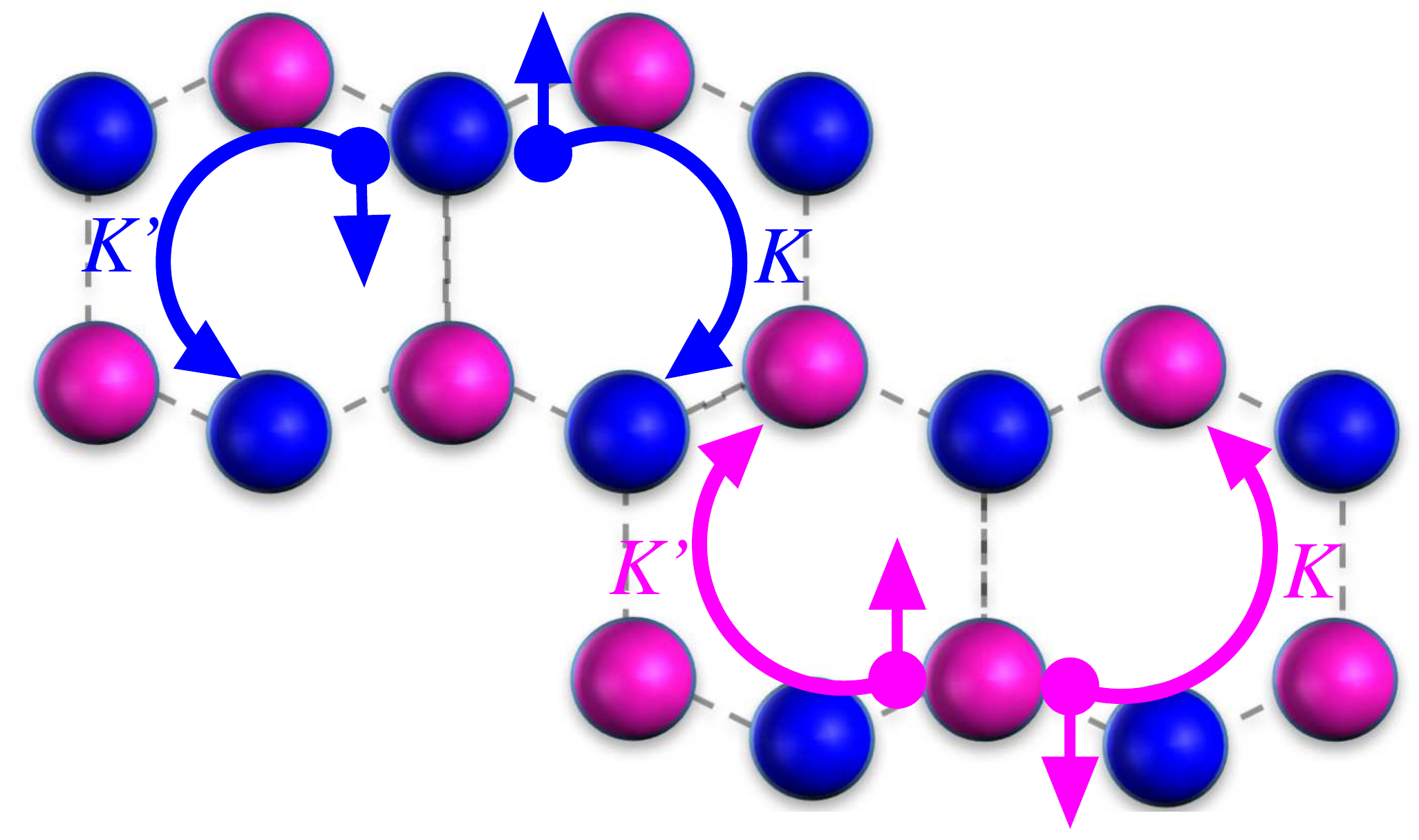}\\
\caption{
Illustration of the spin-valley-orbit coupling of Dirac carriers in the top valence band.
Spin `down'  carriers  couple to counter-clockwise ($m_l=1$) rotating orbitals, 
whereas spin `up' carriers  couple to clockwise ($m_l=-1$) rotating orbitals. 
}
\label{Fig1}
\end{figure}

As pointed out by McClure {\it et al.} \cite{mcclure}, the  
$\pi$-bands are  $p_z$-orbitals hybridized with $d_{xz}$- and $d_{yz}$-orbitals of the nearest neighbor (NN). 
Owing to the large energy difference, the $p_z$-contribution is dominant near the Fermi energy. 
We obtain perturbatively the $d$-band contribution, \cite{konschuh2011spin,konschuh2010tight,huertas2006spin}
\begin{equation}
|\varphi^{(1)}_d\rangle  = 
\frac{3i\tau V_{pd\pi}}{\sqrt{2}\varepsilon_{pd}}
\left[c_A |2 \tau \rangle^B + c_B |2 -\tau \rangle^A\right], 
\label{eqd1}
\end{equation}
where $\varepsilon_{pd}$ is the energetic difference between the $d$- and $p$-orbitals and $V_{pd\pi}$ is the relevant $p-d$ coupling. 
Here, we have expressed the $d$-orbitals in the angular momentum representation, 
$|l,m_l \rangle$, with $\sqrt{2}|2 \pm 1 \rangle =| d_{xz}\rangle \mp i|d_{yz}\rangle $. 
It is worth noting in Eq.(\ref{eqd1}) that $m_l$ relates to the valley index, $\tau$,  commonly termed 
as \emph{valley-orbit} coupling. This is connected to the chirality of the Dirac electrons: in one sublattice, the 
$p_z$-electrons couple to those of  $d$-orbital with  $m_l=1$ ($m_l=-1$) in valley $K$ ($K^\prime$), and the 
converse occurs for the other sublattice \cite{prada}. 
As we will see below, this has important consequences \textcolor{black}{for} the $g$-factor corrections (see diagram of Fig. \ref{Fig1}).

On the other hand, the $\sigma$-band is constituted of $s$ and $p_{x,y}$ of the NN \cite{dresselhaus}. 
The $\pi$-bands described by Eqs.(\ref{eq0}) and (\ref{eqd1}) 
can mix with the $\sigma$-bands 
either  intrinsically  via atomic spin-orbit interaction or 
extrinsically, via structural SOC or Bychkov-Rashba effect.  
The latter emerges as the horizontal mirror symmetry breaks and is 
linear in (uniaxial) electric field   \cite{rashba2009graphene,konschuh2010tight,min2006intrinsic,yao2007spin},
leading to an atomic dipole moment, commonly termed as Stark effect. 
Microscopically, the induced dipole  results in  a non-zero intra-atomic coupling between  the $p_z$- and $s$-orbitals. 
The structural SOC is related to a horizontal plane mirror asymmetry (PIA)
\cite{kochan2,kochan,falko,chengLiu} originated  by ripples, defects or adsorbates, coupling $p_z$ and $p_{x,y}$-orbitals. 

The $\sigma$-band mixing near the Fermi energy is expected to be smaller than the $d$-band contribution, since the 
atomic SOC parameter and the Stark parameter,  $\lambda_z = e E \langle s |\hat z|p_z\rangle $ are small compared to the $p-d$ coupling, 
$\lambda_{\rm soc}^p , \lambda_a, \lambda_z\ll V_{pd\pi}$. We thus consider the $\sigma$-band mixing perturbatively, %
$\hat V  = \hat V_{\rm soc} + \hat V_{\rm PIA} +\hat V_{\rm EF} $, with: 
\begin{eqnarray}
\hat V &=& 
i  \epsilon_{ijk} \hat s_k(\lambda_{\rm{soc}}^p |p_i^\alpha\rangle\langle p_j^\alpha|+
\lambda_{\rm{a}} |p_i^\alpha\rangle\langle p_j^{\overline\alpha}| + 
\lambda_z  |s^\alpha_i\rangle\langle p_z^\alpha|)  +  
\rm{hc} 
\nonumber
\end{eqnarray}
where we have included the atomic $p$-orbital coupling $\lambda_{\rm{soc}}^p$ and structural SOC in $\lambda_a$ and the 
Einstein summation convention is assumed.
The projection over the orbital eigenstates yielding finite angular momentum contributions are \cite{prada}, 
\begin{eqnarray}
\mathcal{\hat P} |\varphi^{(1)}_\sigma\rangle &=& 
(\tau \alpha_{I}^\sigma c_A -is_z\alpha_{E}^\sigma c_B) 
|1-\tau\rangle^A  -\nonumber \\ &&- 
(\tau \alpha_{I}^\sigma c_B -is_z\alpha_{E}^\sigma c_A )
|1\tau\rangle^B,
\label{eqSigma1}
\end{eqnarray}
where 
$\mathcal{\hat  P} = |11\rangle\langle11| + |1-1\rangle\langle1-1|$ 
and $\alpha_{I}^\sigma$, $\alpha_{E}^\sigma$ are the $\sigma$-band intrinsic and extrinsic SOC coefficients:
\[
\alpha_{I}^\sigma = \sqrt{2}\lambda_{\rm soc}^p\left( \frac{\sin^2{\gamma}}{\varepsilon_\sigma^+} 
+ \frac{\cos^2{\gamma}}{\varepsilon_\sigma^-}\right), 
\ \alpha_{E}^\sigma = \alpha_{\rm BR} + \alpha_{\rm PIA} 
\]
with $\tan{\gamma} = 3\sqrt{2}V_{sp\sigma}/2\varepsilon_\sigma^+ $, 
$V_{sp\sigma}$ being the $\sigma$-coupling of the $p$- and $s$-orbitals
and $ \varepsilon_\sigma^{\pm} = \varepsilon_s \pm  \sqrt{ \varepsilon_s^2 + 2(3V_{sp\sigma})^2})/2$. 
Finally,  $\alpha_{\rm BR}$ and $\alpha_{\rm PIA}$ accounts for the Bychkov-Rashba and the  SL asymmetry SOC, respectively.

Eqs.({\ref{eqd1}}) and ({\ref{eqSigma1}}) yield three different second order contributions for $\langle \hat L_z \rangle$,
that is, $\Delta g_{zz} = \sum_i \langle \varphi_i^{(1)} |\hat L_z|\varphi_i^{(1)} \rangle$, giving: 
\begin{equation} 
\Delta g_{zz} \simeq   \tau 
\sigma_z^{0}
\left(
 \left|\frac{\lambda_I}{\lambda_{\rm soc}^d}\right| 
-(\alpha_I^\sigma)^2
+( \alpha_{E}^\sigma)^2
\right),
\label{eqgzz}
\end{equation}
where we have defined $  \sigma_z ^{0} \equiv  \langle \varphi^{(0)}|\hat \sigma_z |\varphi^{(0)}\rangle = |c_A|^2 - |c_B|^2$, 
and we have used the result of Konschuh {\it et al.} \cite{konschuh2010tight}, 
$\lambda_I = 9 V_{pd\pi}^2 \lambda_{\rm soc}^d/(2\epsilon_{pd}^2) $, 
with $\lambda_{\rm soc}^d$ being the atomic SOC for the $d$-orbitals. 
We note that all three terms are proportional to $\tau \langle\sigma_z\rangle$, due to the valley-orbit coupling, and the first 
term is dominant, as we will see.

Fig.\ref{Fig1} illustrates the underlying nature of the spin-valley-orbit coupling for the lowest bands given in Eq.(\ref{eqGS}). 
The highest populated state  is  characterized by $\tau s_z \sigma_z = -1$. 
For sublattice  $B$ (blue, $\sigma_z =-1$), the state has spin `up' in the $K$-valley,  
and it couples to an anti-clockwise rotating $d$-orbital, whereas the spin `down' in the $K^\prime$-valley couples to the clockwise 
rotating $d$ orbital. The converse occurs for sublattice  $A$ (magenta, $\sigma_z = 1$), where  the spin `up' (`down') is in 
the $K^\prime$- ($K$-) valley, but it couples to the $m_l = -1$ ($m_l=1$) $d$-orbital. 

\emph{Hence, in the presence of spin-valley-orbit coupling, the Dirac carrier's spin direction opposes that of the $m_l$ 
quantum number of the coupled $d$-orbital, reducing  the effective $g$-factor at leading order.}

We now consider the in-plane corrections,  $\langle L_{x}\rangle$ and $\langle L_{y}\rangle$, 
with  $\Delta g_{\alpha\alpha} =  2\Re{\langle\varphi_\pm^{(0)} |\hat L_\alpha|\varphi_{\sigma}^{(1)}\rangle}$. 
We choose the $\hat x$ axis to be parallel to a zig-zag direction. 
The theoretical model assumes a well-defined crystalline  zig-zag direction, 
which can be generalized as the transport direction in the polycrystalline, continuum limit.  
Using Eq.(\ref{eqSigma1}) and $\sqrt{2}\hat L_x|1,\pm\tau\rangle^\alpha = |p_z^\alpha\rangle$, 
we obtain first order corrections: 
\begin{eqnarray}
\Delta g_{xx }  &=&
\pm 2\sqrt{2} (\alpha_{I}^\sigma\tau \sigma_z^0 +\alpha_E^\sigma s_z\tau\sin{\varphi_q}),  \nonumber \\
\Delta g_{yy }  &=& \mp  2\sqrt{2}\alpha_E^\sigma s_z \tau\cos{\varphi_q}. 
\label{eqgxy}
\end{eqnarray}
The intrinsic contribution results in  a (dominant) negative  correction for the  highest populated band, 
whereas the sign of the extrinsic one depends on the electric field direction. 
In a single-particle theoretical picture, {\it all} corrections would vanish, as $\langle \tau \hat \sigma_z\rangle$ 
averages out to zero. However, under real experimental conditions and in a macroscopic graphene sample with a spin 
imbalance $n_{\uparrow} - n_{\downarrow} \neq 0$, the problem becomes many-body and corrections to the $g$-factor emerge.

\begin{figure}[!hbt]
{
\includegraphics[angle=0, width=1.\linewidth]{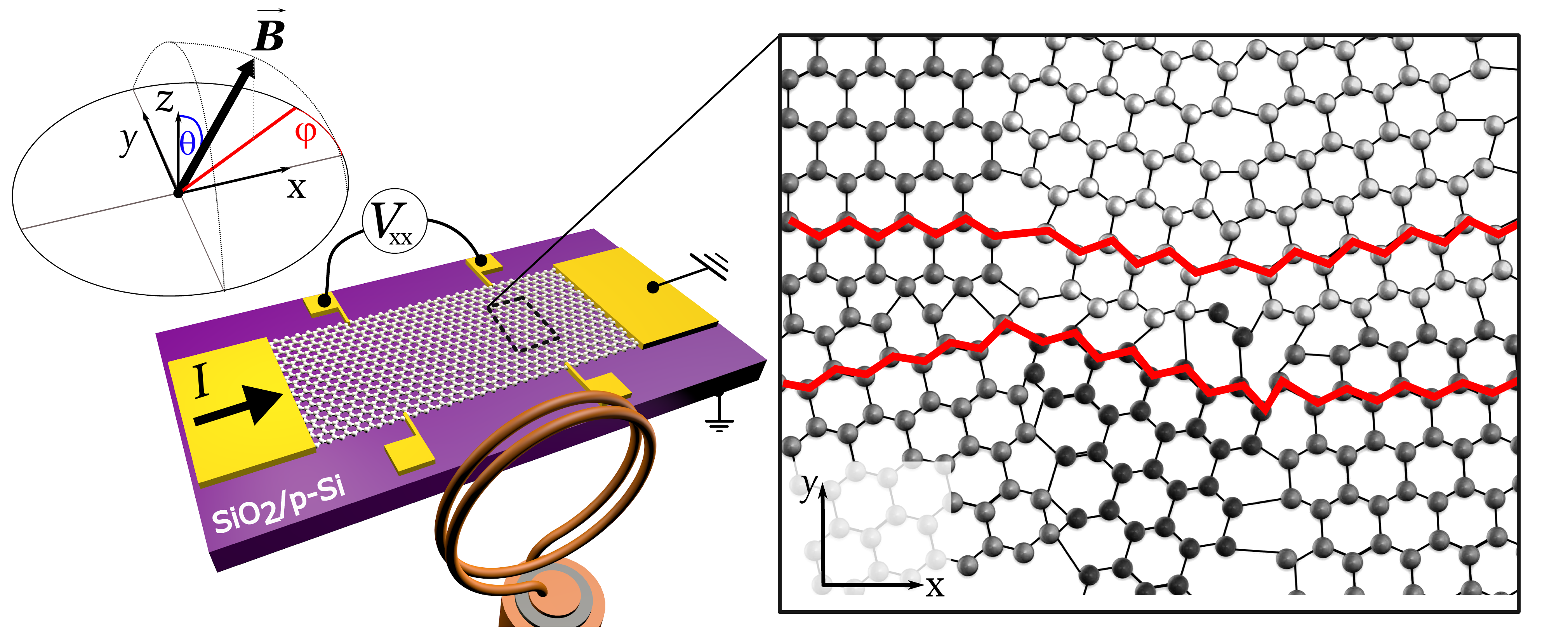}}\\
\caption{  
Schematic setup of our ESR measurements. 
The external magnetic field $|B| \lesssim $ 1 T can freely \textcolor{black}{rotate}, while a 
Hertzian loop antenna induces an AC field. 
A constant current flows along the $x$ direction, and the longitudinal voltage $V_{xx}$ is measured. 
The graphene layer rests on 300nm SiO$_2$ on top of a $p$-Si substrate that is grounded. 
The blow-up on the right-hand side illustrates the granular nature of the CVD graphene. 
The polycrystallinity retains zig-zag directions (red bold lines) parallel to the transport current.
}
\label{Fig2}
\end{figure}

\begin{figure}[!htb]
\includegraphics[angle=0, width=.75\linewidth]{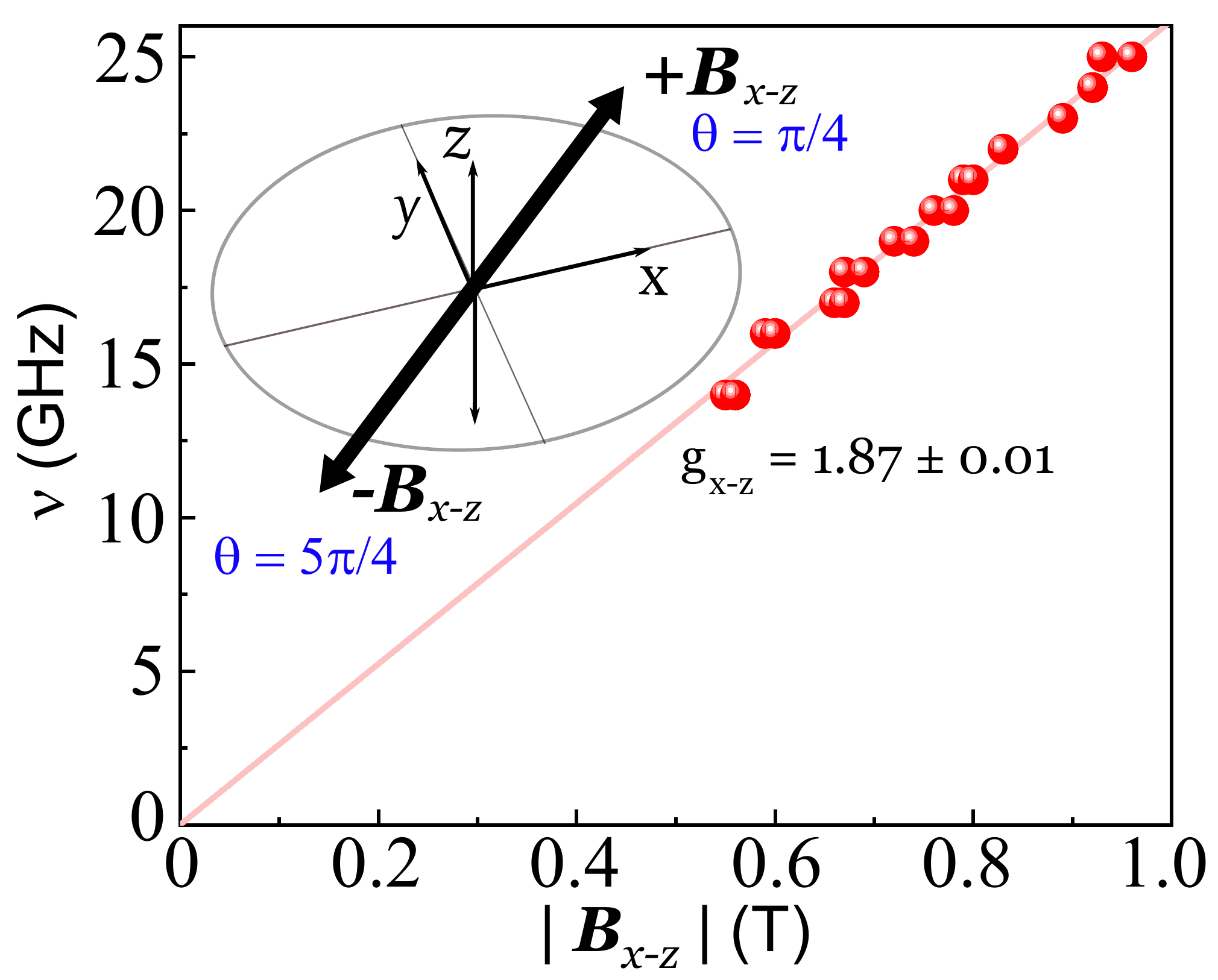}
\caption{(b) Resonance frequency as a function of field strength $|\vec B|$  at constant 
$\theta$ in the $x-z$ plane. The data are taken at $T = 1.4$K, with a microwave radiation power of 21 dBm. The $g$-factor is given by the slope of the linear fit, $g_{xz} = 1.87\pm0.01$.}
\label{Fig3}
\end{figure}

We experimentally scrutinize the spin-valley-orbit coupling and the validity of our model by studying the $g$-tensor in a large-scale (1960 
{$\mu$m} $\times$ 66 $\mu$m) graphene Hall bar on SiO$_2$. 
The device fabrication processes of the graphene that was synthesized by chemical vapor deposition is described by Lyon {\it et al.} \cite{lyon2017upscaling}.  We employed low temperature (1.4K) resistively-detected electron-spin resonance (RD-ESR) \cite{jonas,lyon2017probing,mani2012observation}, 
a spin-selective probing technique that couples carriers of opposite spin by microwave excitation, and detects the response resistively. 
The large dimension of our device ensures the continuum limit with a well-defined bulk gap and chirality for the charge carriers \cite{kane2005quantum}, and the polycrystalline nature of the sample retains the theorized zig-zag directions parallel to the transport directions, as illustrated in Fig.\ref{Fig2}. Polycrystallinity also induces disorder, which broadens the resonant signal and facilitates the resistive detection.

The microwave excitation field is generated by a Hertzian loop antenna adjacent to the sample [see Fig. \ref{Fig2}]. 
A constant low frequency current $I$ = 1nA is passed through the sample along $x$-direction, 
which we can relate to the \emph{propagating} zig-zag direction [blow-up in Fig. \ref{Fig2}], while a standard lock-in technique probes the resulting longitudinal resistance, $R_{xx}= V_{xx} / I$, as a function of 
the magnetic field $\vec B$. The magnetic field vector $\vec B$ can freely rotate with respect to the sample plane. Hence,  
we use spherical coordinates to denote the orientation of $\vec B$, with $\theta$ as the out-of-plane angle and $\varphi$ for 
in-plane rotations. All measurement are performed without the application of a gate voltage (substrate grounded), corresponding to 
an intrinsic density of $n \approx 6\times 10^{11}$ cm$^{-2}$. 

Whenever the microwave frequency $\nu$ matches the resonant condition of Eq.(\ref{eqR}), 
the increased band population reduces $R_{xx,{\nu}}$. 
We can resolve these resonances as peaks in the microwave-induced differential resistance, $\Delta R_{xx}(\nu)=R_{xx,{\rm dark}} - R_{xx,{\nu}}$. 

\begin{center}
\begin{figure*}[!htb]
\includegraphics[angle=0, width=1.0\linewidth]{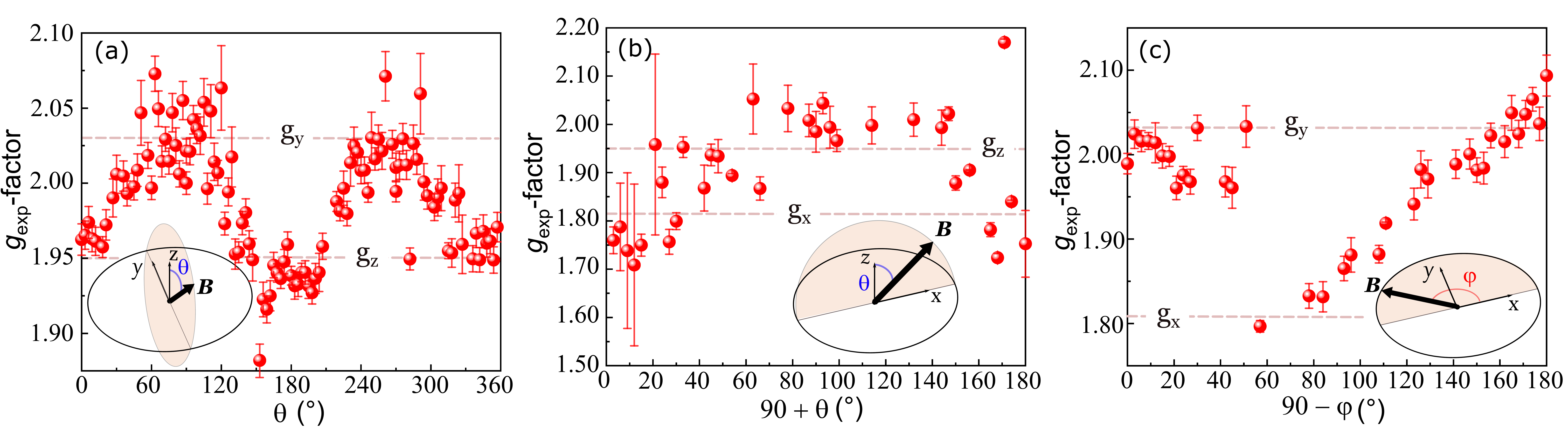}
\caption{
 Angular dependence of the $g$-factor: (a) $g$-factor for a rotation of $\theta$ in the $y-z$ plane and (b) for a rotation of $\theta$ in the $x-y$ plane and (c) for a rotation of $\varphi$ in the $x-y$ plane. Along the in-plane, we extract $\Delta g_{xx} = -0.19\pm 0.01$ and $\Delta g_{yy} = 0.03\pm 0.01$, whereas $\Delta g_{zz} = -0.05\pm 0.01$. 
}
\label{Fig4}
\end{figure*}
\end{center}

Fig. \ref{Fig3} shows the electron spin resonances for different values of $|\vec B|$ under constant angle $\theta$ in the $x-z$ plane. Each data point is the result of a Gaussian fit to the resonance curves $\Delta R_{xx}(\nu)$ (not shown). The data points follow a linear dispersion that reflects the magnetic field dependence of the Zeeman energy. Its slope thus represents the \emph{effective} $g$-factor of the Dirac electrons for the magnetic field under $\theta$.

The procedure is repeated for $\vec B$ pointing in different directions, i.e. for various $\theta$ and $\varphi$, allowing us to resolve the anisotropic $g$-factor as defined in Eq.(\ref{eqgeff}). Fig.\ref{Fig4} is an angle-resolved study of the effective $g$-factor within the planes marked schematically inside each graph. 
In Fig.\ref{Fig4}(a) we explore the $y-z$ plane, where a sinusoidal dependence of the $g$-factor on the axial angle $\theta$ is apparent. 
When the external field is oriented perpendicular to the sample plane ($\theta =0$), we obtain $g_{zz} = 1.95\pm 0.02$\cite{jonas}, whereas for $\theta = 90^o$, we obtain $g_{yy} = 2.03\pm 0.02$. Fig.\ref{Fig4}(b) shows a rotation of $\theta$  in the 
$y=0$ plane. Here, the effective $g$-factor is smallest when $\vec B$ is collinear to the current direction and becomes $g_{xx} = 1.81\pm 0.02$.  The in-plane variation of $\varphi$ at fixed $\theta = \pi/2$ is shown in Fig.\ref{Fig4}(c) for completeness. 
We attribute the asymmetry to the current-induced Rashba effect \cite{wilamowski} and to the polycrystalline structure of the graphene, where the current direction is defined only locally, as illustrated in Fig. \ref{Fig2}. 
The largest  correction is thus obtained for $\Delta g_{xx} = -0.19\pm 0.01$,
consistent with the first-order intrinsic SOC, and the smallest correction is 
$\Delta g_{yy} = 0.03\pm 0.01$, consistent with a small extrinsic SOC in the absence of gating. 
Finally,  $\Delta g_{zz} = -0.05\pm 0.01$, corresponding to a second order correction.
Table \ref{tab1} summarizes the experimentally extracted elements of the $g$-tensor and the $g$-factor anisotropy.
\begin{table}[hbt!]
\caption{\label{tab1}
\textcolor{black}{Experimentally determined effective} $g$-factors for the \textcolor{black}{three orthogonal directions} obtained at 1.4 K. 
}
\begin{ruledtabular}
\begin{tabular}{lllllll}
$g_x$ & $g_{x-y}$ & $g_{y}$ & $g_{y-z}$ & $g_z$ & $g_{xz}$ &  Error \\
\hline \\
 1.81 & 1.91 & 2.03 & 1.99 & 1.95 & 1.87 & 0.01 \\
\end{tabular}
\end{ruledtabular}
\end{table}

We can extract  atomic SOC parameters that lead to the observed $g$-factor corrections. 
For the in-plane corrections,  using (\ref{eqgxy}) with $\varphi_q=0$,  
we obtain $\alpha_E^\sigma \simeq (0.01\pm 0.03)$. 
We also obtain 
$\alpha_{I}^\sigma  \simeq ( 0.067 \pm 0.003)$, 
giving an upper limit for $\lambda_{\rm soc}^p < 0.05\varepsilon_s$, consistent with the theoretical value
\cite{cardona,hermann,min2006intrinsic,yao2007spin,huertas2006spin,konschuh2010tight, dresselhaus}. 

For the axial correction, using $\lambda_I\simeq$ 21 $\mu$eV \cite{jonas} in Eq. (\ref{eqgzz}) %
we obtain SOC parameter for $d$-orbitals, $\lambda_{\rm soc}^d$, 
\[
\lambda_{\rm soc}^d \simeq (0.31 \pm 0.09) {\rm \ meV},
\]
which compares quite well with the DFT obtained value of 0.8 meV  \cite{konschuh2010tight}. 
As pointed \textcolor{black}{out} by Konschuh {\it et al.}, unlike in the $\lambda_{\rm soc}^p$ case, 
there is no possible fitting of the energy spectrum  to obtain numerically this value, 
since the needed high-energy states in the conduction bands cannot be identified. 
We stress that this value is {\it intrinsic}: it is the atomic spin-orbit coupling 
not only for \textcolor{black}{the special case of} graphene, but for all $^{12}$C atoms.

In summary, we experimentally resolved the $g$-factor anisotropy in graphene using an angle-dependent ESR method. 
The chiral nature of the Dirac electrons in graphene entails corrections to the $g$-factor that originate 
from a peculiar spin-valley orbit coupling. 
Along the transport direction, we observe a negative first order correction, owing to the intrinsic, chiral SOC 
with the propagating $p_x$-orbital. We extract an intrinsic coupling of $\alpha_I^\sigma \simeq 0.067\pm 0.003$. 
Along the $y$-direction, the sign and magnitude of the $g$-factor correction reflects a extrinsic SOC, consistent with 
the absence of inversion symmetry. We extract an extrinsic coupling of $\alpha_E^\sigma \simeq 0.001\pm 0.003$. 
In combination with the axial correction, we were able to extract intrinsic SOC parameter 
$\lambda_{\rm soc}^d = 0.31 \pm 0.09 $ meV. 

We acknowledge support by the Bundesminsterium f\"ur Forschung und Technologie (BMBF) through the 
`Forschungslabor Mikroelectronik Deutschland (ForLab)'. We thank Hans-Peter Oepen and T. Schmirander for fruitful discussions.
All experiments were performed with nanomeas (www.nanomeas.com).

\bibliography{bibliography}

\providecommand{\noopsort}[1]{}\providecommand{\singleletter}[1]{#1}%
\begin{thebibliography}{40}%
\makeatletter
\providecommand \@ifxundefined [1]{%
 \@ifx{#1\undefined}
}%
\providecommand \@ifnum [1]{%
 \ifnum #1\expandafter \@firstoftwo
 \else \expandafter \@secondoftwo
 \fi
}%
\providecommand \@ifx [1]{%
 \ifx #1\expandafter \@firstoftwo
 \else \expandafter \@secondoftwo
 \fi
}%
\providecommand \natexlab [1]{#1}%
\providecommand \enquote  [1]{``#1''}%
\providecommand \bibnamefont  [1]{#1}%
\providecommand \bibfnamefont [1]{#1}%
\providecommand \citenamefont [1]{#1}%
\providecommand \href@noop [0]{\@secondoftwo}%
\providecommand \href [0]{\begingroup \@sanitize@url \@href}%
\providecommand \@href[1]{\@@startlink{#1}\@@href}%
\providecommand \@@href[1]{\endgroup#1\@@endlink}%
\providecommand \@sanitize@url [0]{\catcode `\\12\catcode `\$12\catcode
  `\&12\catcode `\#12\catcode `\^12\catcode `\_12\catcode `\%12\relax}%
\providecommand \@@startlink[1]{}%
\providecommand \@@endlink[0]{}%
\providecommand \url  [0]{\begingroup\@sanitize@url \@url }%
\providecommand \@url [1]{\endgroup\@href {#1}{\urlprefix }}%
\providecommand \urlprefix  [0]{URL }%
\providecommand \Eprint [0]{\href }%
\providecommand \doibase [0]{http://dx.doi.org/}%
\providecommand \selectlanguage [0]{\@gobble}%
\providecommand \bibinfo  [0]{\@secondoftwo}%
\providecommand \bibfield  [0]{\@secondoftwo}%
\providecommand \translation [1]{[#1]}%
\providecommand \BibitemOpen [0]{}%
\providecommand \bibitemStop [0]{}%
\providecommand \bibitemNoStop [0]{.\EOS\space}%
\providecommand \EOS [0]{\spacefactor3000\relax}%
\providecommand \BibitemShut  [1]{\csname bibitem#1\endcsname}%
\let\auto@bib@innerbib\@empty
\bibitem [{\citenamefont {Dirac}(1928)}]{dirac1928}%
  \BibitemOpen
  \bibfield  {author} {\bibinfo {author} {\bibfnamefont {P.~A.~M.}\
  \bibnamefont {Dirac}},\ }\href {http://www.jstor.org/stable/94981} {\bibfield
   {journal} {\bibinfo  {journal} {Proceedings of the Royal Society of London.
  Series A, Containing Papers of a Mathematical and Physical Character}\
  }\textbf {\bibinfo {volume} {117}},\ \bibinfo {pages} {610} (\bibinfo {year}
  {1928})}\BibitemShut {NoStop}%
\bibitem [{\citenamefont {Greiner}(1994)}]{greiner}%
  \BibitemOpen
  \bibfield  {author} {\bibinfo {author} {\bibfnamefont {W.}~\bibnamefont
  {Greiner}},\ }\href@noop {} {\emph {\bibinfo {title} {Relativistic Quantum
  Mechanics}}}\ (\bibinfo  {publisher} {Springer},\ \bibinfo {year}
  {1994})\BibitemShut {NoStop}%
\bibitem [{\citenamefont {Novoselov}\ \emph {et~al.}(2005)\citenamefont
  {Novoselov}, \citenamefont {Geim}, \citenamefont {Morozov}, \citenamefont
  {Jiang}, \citenamefont {Katsnelson}, \citenamefont {Grigorieva},
  \citenamefont {Dubonos},\ and\ \citenamefont {Firsov}}]{novoselov2005two}%
  \BibitemOpen
  \bibfield  {author} {\bibinfo {author} {\bibfnamefont {K.~S.}\ \bibnamefont
  {Novoselov}}, \bibinfo {author} {\bibfnamefont {A.~K.}\ \bibnamefont {Geim}},
  \bibinfo {author} {\bibfnamefont {S.~V.}\ \bibnamefont {Morozov}}, \bibinfo
  {author} {\bibfnamefont {D.}~\bibnamefont {Jiang}}, \bibinfo {author}
  {\bibfnamefont {M.~I.}\ \bibnamefont {Katsnelson}}, \bibinfo {author}
  {\bibfnamefont {I.~V.}\ \bibnamefont {Grigorieva}}, \bibinfo {author}
  {\bibfnamefont {S.~V.}\ \bibnamefont {Dubonos}}, \ and\ \bibinfo {author}
  {\bibfnamefont {A.~A.}\ \bibnamefont {Firsov}},\ }\href@noop {} {\bibfield
  {journal} {\bibinfo  {journal} {Nature}\ }\textbf {\bibinfo {volume} {438}},\
  \bibinfo {pages} {197} (\bibinfo {year} {2005})}\BibitemShut {NoStop}%
\bibitem [{\citenamefont {Katsnelson}\ \emph {et~al.}(2006)\citenamefont
  {Katsnelson}, \citenamefont {Novoselov},\ and\ \citenamefont
  {Geim}}]{katsnelson2006chiral}%
  \BibitemOpen
  \bibfield  {author} {\bibinfo {author} {\bibfnamefont {M.~I.}\ \bibnamefont
  {Katsnelson}}, \bibinfo {author} {\bibfnamefont {K.}~\bibnamefont
  {Novoselov}}, \ and\ \bibinfo {author} {\bibfnamefont {A.~K.}\ \bibnamefont
  {Geim}},\ }\href@noop {} {\bibfield  {journal} {\bibinfo  {journal} {Nature
  Physics}\ }\textbf {\bibinfo {volume} {2}},\ \bibinfo {pages} {620} (\bibinfo
  {year} {2006})}\BibitemShut {NoStop}%
\bibitem [{\citenamefont {Geim}\ and\ \citenamefont
  {Novoselov}(2010)}]{geim2010rise}%
  \BibitemOpen
  \bibfield  {author} {\bibinfo {author} {\bibfnamefont {A.~K.}\ \bibnamefont
  {Geim}}\ and\ \bibinfo {author} {\bibfnamefont {K.~S.}\ \bibnamefont
  {Novoselov}},\ }in\ \href@noop {} {\emph {\bibinfo {booktitle} {Nanoscience
  and Technology: A Collection of Reviews from Nature Journals}}}\ (\bibinfo
  {publisher} {World Scientific},\ \bibinfo {year} {2010})\ pp.\ \bibinfo
  {pages} {11--19}\BibitemShut {NoStop}%
\bibitem [{\citenamefont {Nair}\ \emph {et~al.}(2008)\citenamefont {Nair},
  \citenamefont {Blake}, \citenamefont {Grigorenko}, \citenamefont {Novoselov},
  \citenamefont {Booth}, \citenamefont {Stauber}, \citenamefont {Peres},\ and\
  \citenamefont {Geim}}]{Nair08}%
  \BibitemOpen
  \bibfield  {author} {\bibinfo {author} {\bibfnamefont {R.~R.}\ \bibnamefont
  {Nair}}, \bibinfo {author} {\bibfnamefont {P.}~\bibnamefont {Blake}},
  \bibinfo {author} {\bibfnamefont {A.~N.}\ \bibnamefont {Grigorenko}},
  \bibinfo {author} {\bibfnamefont {K.~S.}\ \bibnamefont {Novoselov}}, \bibinfo
  {author} {\bibfnamefont {T.~J.}\ \bibnamefont {Booth}}, \bibinfo {author}
  {\bibfnamefont {T.}~\bibnamefont {Stauber}}, \bibinfo {author} {\bibfnamefont
  {N.~M.~R.}\ \bibnamefont {Peres}}, \ and\ \bibinfo {author} {\bibfnamefont
  {A.~K.}\ \bibnamefont {Geim}},\ }\href {\doibase 10.1126/science.1156965} {\
  \textbf {\bibinfo {volume} {320}},\ \bibinfo {pages} {1308} (\bibinfo {year}
  {2008})}\BibitemShut {NoStop}%
\bibitem [{\citenamefont {Mecklenburg}\ and\ \citenamefont
  {Regan}(2011)}]{regan}%
  \BibitemOpen
  \bibfield  {author} {\bibinfo {author} {\bibfnamefont {M.}~\bibnamefont
  {Mecklenburg}}\ and\ \bibinfo {author} {\bibfnamefont {B.~C.}\ \bibnamefont
  {Regan}},\ }\href {\doibase 10.1103/PhysRevLett.106.116803} {\bibfield
  {journal} {\bibinfo  {journal} {Phys. Rev. Lett.}\ }\textbf {\bibinfo
  {volume} {106}},\ \bibinfo {pages} {116803} (\bibinfo {year}
  {2011})}\BibitemShut {NoStop}%
\bibitem [{\citenamefont {Golub}\ \emph {et~al.}(2020)\citenamefont {Golub},
  \citenamefont {Egger}, \citenamefont {M\"uller},\ and\ \citenamefont
  {Villalba-Ch\'avez}}]{golub}%
  \BibitemOpen
  \bibfield  {author} {\bibinfo {author} {\bibfnamefont {A.}~\bibnamefont
  {Golub}}, \bibinfo {author} {\bibfnamefont {R.}~\bibnamefont {Egger}},
  \bibinfo {author} {\bibfnamefont {C.}~\bibnamefont {M\"uller}}, \ and\
  \bibinfo {author} {\bibfnamefont {S.}~\bibnamefont {Villalba-Ch\'avez}},\
  }\href {\doibase 10.1103/PhysRevLett.124.110403} {\bibfield  {journal}
  {\bibinfo  {journal} {Phys. Rev. Lett.}\ }\textbf {\bibinfo {volume} {124}},\
  \bibinfo {pages} {110403} (\bibinfo {year} {2020})}\BibitemShut {NoStop}%
\bibitem [{\citenamefont {Giuliani}\ \emph {et~al.}(2012)\citenamefont
  {Giuliani}, \citenamefont {Mastropietro},\ and\ \citenamefont
  {Porta}}]{GIULIANI2012461}%
  \BibitemOpen
  \bibfield  {author} {\bibinfo {author} {\bibfnamefont {A.}~\bibnamefont
  {Giuliani}}, \bibinfo {author} {\bibfnamefont {V.}~\bibnamefont
  {Mastropietro}}, \ and\ \bibinfo {author} {\bibfnamefont {M.}~\bibnamefont
  {Porta}},\ }\href {\doibase https://doi.org/10.1016/j.aop.2011.10.007}
  {\bibfield  {journal} {\bibinfo  {journal} {Annals of Physics}\ }\textbf
  {\bibinfo {volume} {327}},\ \bibinfo {pages} {461 } (\bibinfo {year}
  {2012})}\BibitemShut {NoStop}%
\bibitem [{\citenamefont {Kane}\ and\ \citenamefont
  {Mele}(2005{\natexlab{a}})}]{kane2005quantum}%
  \BibitemOpen
  \bibfield  {author} {\bibinfo {author} {\bibfnamefont {C.~L.}\ \bibnamefont
  {Kane}}\ and\ \bibinfo {author} {\bibfnamefont {E.~J.}\ \bibnamefont
  {Mele}},\ }\href@noop {} {\bibfield  {journal} {\bibinfo  {journal} {Physical
  Review Letters}\ }\textbf {\bibinfo {volume} {95}},\ \bibinfo {pages}
  {226801} (\bibinfo {year} {2005}{\natexlab{a}})}\BibitemShut {NoStop}%
\bibitem [{\citenamefont {Kane}\ and\ \citenamefont
  {Mele}(2005{\natexlab{b}})}]{kane2005z}%
  \BibitemOpen
  \bibfield  {author} {\bibinfo {author} {\bibfnamefont {C.~L.}\ \bibnamefont
  {Kane}}\ and\ \bibinfo {author} {\bibfnamefont {E.~J.}\ \bibnamefont
  {Mele}},\ }\href@noop {} {\bibfield  {journal} {\bibinfo  {journal} {Physical
  Review Letters}\ }\textbf {\bibinfo {volume} {95}},\ \bibinfo {pages}
  {146802} (\bibinfo {year} {2005}{\natexlab{b}})}\BibitemShut {NoStop}%
\bibitem [{\citenamefont {Sichau}\ \emph {et~al.}(2019)\citenamefont {Sichau},
  \citenamefont {Prada}, \citenamefont {Anlauf}, \citenamefont {Lyon},
  \citenamefont {Bosnjak},\ and\ \citenamefont {Tiemann}}]{jonas}%
  \BibitemOpen
  \bibfield  {author} {\bibinfo {author} {\bibfnamefont {J.}~\bibnamefont
  {Sichau}}, \bibinfo {author} {\bibfnamefont {M.}~\bibnamefont {Prada}},
  \bibinfo {author} {\bibfnamefont {T.}~\bibnamefont {Anlauf}}, \bibinfo
  {author} {\bibfnamefont {T.~J.}\ \bibnamefont {Lyon}}, \bibinfo {author}
  {\bibfnamefont {B.}~\bibnamefont {Bosnjak}}, \ and\ \bibinfo {author}
  {\bibfnamefont {R.}~\bibnamefont {Tiemann}, \bibfnamefont {L.~Blick}},\
  }\href@noop {} {\bibfield  {journal} {\bibinfo  {journal} {Physical Review
  Letters}\ }\textbf {\bibinfo {volume} {122}},\ \bibinfo {pages} {046403}
  (\bibinfo {year} {2019})}\BibitemShut {NoStop}%
\bibitem [{\citenamefont {Prada}()}]{prada}%
  \BibitemOpen
  \bibfield  {author} {\bibinfo {author} {\bibfnamefont {M.}~\bibnamefont
  {Prada}},\ }\href@noop {} {\ }\BibitemShut {NoStop}%
\bibitem [{\citenamefont {Weil}\ and\ \citenamefont {Bolton}(2006)}]{weil}%
  \BibitemOpen
  \bibfield  {author} {\bibinfo {author} {\bibfnamefont {J.~A.}\ \bibnamefont
  {Weil}}\ and\ \bibinfo {author} {\bibfnamefont {J.~R.}\ \bibnamefont
  {Bolton}},\ }\href
  {https://onlinelibrary.wiley.com/doi/abs/10.1002/9780470084984} {\emph
  {\bibinfo {title} {Electron Paramagnetic Resonance: Elementary Theory and
  Practical Applications}}}\ (\bibinfo  {publisher} {John Wiley \& Sons, Ltd},\
  \bibinfo {year} {2006})\BibitemShut {NoStop}%
\bibitem [{\citenamefont {Bloch}(1958)}]{bloch}%
  \BibitemOpen
  \bibfield  {author} {\bibinfo {author} {\bibfnamefont {C.}~\bibnamefont
  {Bloch}},\ }\href@noop {} {\bibfield  {journal} {\bibinfo  {journal} {Nucl.
  Phys}\ }\textbf {\bibinfo {volume} {6}},\ \bibinfo {pages} {329} (\bibinfo
  {year} {1958})}\BibitemShut {NoStop}%
\bibitem [{\citenamefont {Mostafanejad}()}]{mostafa}%
  \BibitemOpen
  \bibfield  {author} {\bibinfo {author} {\bibfnamefont {M.}~\bibnamefont
  {Mostafanejad}},\ }\href {\doibase https://doi.org/10.1002/qua.24721}
  {\bibfield  {journal} {\bibinfo  {journal} {International Journal of Quantum
  Chemistry}\ }\textbf {\bibinfo {volume} {114}},\ \bibinfo {pages}
  {1495}}\BibitemShut {NoStop}%
\bibitem [{\citenamefont {Rudowicz}(1987)}]{rudo1}%
  \BibitemOpen
  \bibfield  {author} {\bibinfo {author} {\bibfnamefont {C.}~\bibnamefont
  {Rudowicz}},\ }\href {https://books.google.de/books?id=HV8yQwAACAAJ} {\emph
  {\bibinfo {title} {Concept of Spin Hamiltonian, Forms of Zero Field Splitting
  and Electronic Zeeman Hamiltonians and Relations Between Parameters Used in
  EPR: A Critical Review}}},\ Magnetic resonance review\ (\bibinfo {year}
  {1987})\BibitemShut {NoStop}%
\bibitem [{\citenamefont {Slichter}(1990)}]{Slichter1990}%
  \BibitemOpen
  \bibfield  {author} {\bibinfo {author} {\bibfnamefont {C.~P.}\ \bibnamefont
  {Slichter}},\ }\href {\doibase 10.1007/978-3-662-09441-9} {\emph {\bibinfo
  {title} {{Principles of Magnetic Resonance}}}},\ \bibinfo {series} {Springer
  Series in Solid-State Sciences}, Vol.~\bibinfo {volume} {1}\ (\bibinfo
  {publisher} {Springer Berlin Heidelberg},\ \bibinfo {address} {Berlin,
  Heidelberg},\ \bibinfo {year} {1990})\BibitemShut {NoStop}%
\bibitem [{\citenamefont {Neto}\ \emph {et~al.}(2009)\citenamefont {Neto},
  \citenamefont {Guinea}, \citenamefont {Peres}, \citenamefont {Novoselov},\
  and\ \citenamefont {Geim}}]{neto2009electronic}%
  \BibitemOpen
  \bibfield  {author} {\bibinfo {author} {\bibfnamefont {A.~H.~C.}\
  \bibnamefont {Neto}}, \bibinfo {author} {\bibfnamefont {F.}~\bibnamefont
  {Guinea}}, \bibinfo {author} {\bibfnamefont {N.~M.~R.}\ \bibnamefont
  {Peres}}, \bibinfo {author} {\bibfnamefont {K.~S.}\ \bibnamefont
  {Novoselov}}, \ and\ \bibinfo {author} {\bibfnamefont {A.~K.}\ \bibnamefont
  {Geim}},\ }\href@noop {} {\bibfield  {journal} {\bibinfo  {journal} {Reviews
  of Modern Physics}\ }\textbf {\bibinfo {volume} {81}},\ \bibinfo {pages}
  {109} (\bibinfo {year} {2009})}\BibitemShut {NoStop}%
\bibitem [{\citenamefont {Katsnelson}(2012)}]{katsnelson2012graphene}%
  \BibitemOpen
  \bibfield  {author} {\bibinfo {author} {\bibfnamefont {M.~I.}\ \bibnamefont
  {Katsnelson}},\ }\href@noop {} {\emph {\bibinfo {title} {Graphene: Carbon in
  Two Dimensions}}}\ (\bibinfo  {publisher} {Cambridge university press},\
  \bibinfo {year} {2012})\BibitemShut {NoStop}%
\bibitem [{\citenamefont {Singh}\ \emph {et~al.}(2020)\citenamefont {Singh},
  \citenamefont {Prada}, \citenamefont {Strenzke}, \citenamefont {Bosnjak},
  \citenamefont {Schmirander}, \citenamefont {Tiemann},\ and\ \citenamefont
  {Blick}}]{singh2020sublattice}%
  \BibitemOpen
  \bibfield  {author} {\bibinfo {author} {\bibfnamefont {U.~R.}\ \bibnamefont
  {Singh}}, \bibinfo {author} {\bibfnamefont {M.}~\bibnamefont {Prada}},
  \bibinfo {author} {\bibfnamefont {V.}~\bibnamefont {Strenzke}}, \bibinfo
  {author} {\bibfnamefont {B.}~\bibnamefont {Bosnjak}}, \bibinfo {author}
  {\bibfnamefont {T.}~\bibnamefont {Schmirander}}, \bibinfo {author}
  {\bibfnamefont {L.}~\bibnamefont {Tiemann}}, \ and\ \bibinfo {author}
  {\bibfnamefont {R.~H.}\ \bibnamefont {Blick}},\ }\href@noop {} {\bibfield
  {journal} {\bibinfo  {journal} {Phys. Rev. B}\ ,\ \bibinfo {pages} {in
  press}} (\bibinfo {year} {2020})},\ \Eprint {http://arxiv.org/abs/2006.04190}
  {arXiv:2006.04190 [cond-mat.mes-hall]} \BibitemShut {NoStop}%
\bibitem [{\citenamefont {Banszerus}\ \emph {et~al.}(2020)\citenamefont
  {Banszerus}, \citenamefont {Frohn}, \citenamefont {Fabian}, \citenamefont
  {Somanchi}, \citenamefont {Epping}, \citenamefont {M\"uller}, \citenamefont
  {Neumaier}, \citenamefont {Watanabe}, \citenamefont {Taniguchi},
  \citenamefont {Libisch}, \citenamefont {Beschoten}, \citenamefont {Hassler},\
  and\ \citenamefont {Stampfer}}]{BanszerusPRL20}%
  \BibitemOpen
  \bibfield  {author} {\bibinfo {author} {\bibfnamefont {L.}~\bibnamefont
  {Banszerus}}, \bibinfo {author} {\bibfnamefont {B.}~\bibnamefont {Frohn}},
  \bibinfo {author} {\bibfnamefont {T.}~\bibnamefont {Fabian}}, \bibinfo
  {author} {\bibfnamefont {S.}~\bibnamefont {Somanchi}}, \bibinfo {author}
  {\bibfnamefont {A.}~\bibnamefont {Epping}}, \bibinfo {author} {\bibfnamefont
  {M.}~\bibnamefont {M\"uller}}, \bibinfo {author} {\bibfnamefont
  {D.}~\bibnamefont {Neumaier}}, \bibinfo {author} {\bibfnamefont
  {K.}~\bibnamefont {Watanabe}}, \bibinfo {author} {\bibfnamefont
  {T.}~\bibnamefont {Taniguchi}}, \bibinfo {author} {\bibfnamefont
  {F.}~\bibnamefont {Libisch}}, \bibinfo {author} {\bibfnamefont
  {B.}~\bibnamefont {Beschoten}}, \bibinfo {author} {\bibfnamefont
  {F.}~\bibnamefont {Hassler}}, \ and\ \bibinfo {author} {\bibfnamefont
  {C.}~\bibnamefont {Stampfer}},\ }\href {\doibase
  10.1103/PhysRevLett.124.177701} {\bibfield  {journal} {\bibinfo  {journal}
  {Phys. Rev. Lett.}\ }\textbf {\bibinfo {volume} {124}},\ \bibinfo {pages}
  {177701} (\bibinfo {year} {2020})}\BibitemShut {NoStop}%
\bibitem [{\citenamefont {W.}\ and\ \citenamefont {Y.~Yafet}(1962)}]{mcclure}%
  \BibitemOpen
  \bibfield  {author} {\bibinfo {author} {\bibfnamefont {M.~J.}\ \bibnamefont
  {W.}}\ and\ \bibinfo {author} {\bibfnamefont {Y.}~\bibnamefont {Y.~Yafet}},\
  }\href@noop {} {\emph {\bibinfo {title} {Proceedings of the Fifth Conference
  on Carbon}}}\ (\bibinfo  {publisher} {Pergamon, New York},\ \bibinfo {year}
  {1962})\BibitemShut {NoStop}%
\bibitem [{\citenamefont {Konschuh}(2011)}]{konschuh2011spin}%
  \BibitemOpen
  \bibfield  {author} {\bibinfo {author} {\bibfnamefont {S.}~\bibnamefont
  {Konschuh}},\ }\href {https://books.google.de/books?id=MLFgtwAACAAJ} {\emph
  {\bibinfo {title} {Spin-orbit Coupling Effects from Graphene to Graphite}}}\
  (\bibinfo  {publisher} {Universit{\"a}tsbibliothek Regensburg},\ \bibinfo
  {year} {2011})\BibitemShut {NoStop}%
\bibitem [{\citenamefont {Konschuh}\ \emph {et~al.}(2010)\citenamefont
  {Konschuh}, \citenamefont {Gmitra},\ and\ \citenamefont
  {Fabian}}]{konschuh2010tight}%
  \BibitemOpen
  \bibfield  {author} {\bibinfo {author} {\bibfnamefont {S.}~\bibnamefont
  {Konschuh}}, \bibinfo {author} {\bibfnamefont {M.}~\bibnamefont {Gmitra}}, \
  and\ \bibinfo {author} {\bibfnamefont {J.}~\bibnamefont {Fabian}},\
  }\href@noop {} {\bibfield  {journal} {\bibinfo  {journal} {Physical Review
  B}\ }\textbf {\bibinfo {volume} {82}},\ \bibinfo {pages} {245412} (\bibinfo
  {year} {2010})}\BibitemShut {NoStop}%
\bibitem [{\citenamefont {Huertas-Hernando}\ \emph {et~al.}(2006)\citenamefont
  {Huertas-Hernando}, \citenamefont {Guinea},\ and\ \citenamefont
  {Brataas}}]{huertas2006spin}%
  \BibitemOpen
  \bibfield  {author} {\bibinfo {author} {\bibfnamefont {D.}~\bibnamefont
  {Huertas-Hernando}}, \bibinfo {author} {\bibfnamefont {F.}~\bibnamefont
  {Guinea}}, \ and\ \bibinfo {author} {\bibfnamefont {A.}~\bibnamefont
  {Brataas}},\ }\href@noop {} {\bibfield  {journal} {\bibinfo  {journal}
  {Physical Review B}\ }\textbf {\bibinfo {volume} {74}},\ \bibinfo {pages}
  {155426} (\bibinfo {year} {2006})}\BibitemShut {NoStop}%
\bibitem [{\citenamefont {Saito}\ \emph {et~al.}(1998)\citenamefont {Saito},
  \citenamefont {Dresselhaus},\ and\ \citenamefont
  {Dresselhaus}}]{dresselhaus}%
  \BibitemOpen
  \bibfield  {author} {\bibinfo {author} {\bibfnamefont {R.}~\bibnamefont
  {Saito}}, \bibinfo {author} {\bibfnamefont {G.}~\bibnamefont {Dresselhaus}},
  \ and\ \bibinfo {author} {\bibfnamefont {M.~S.}\ \bibnamefont
  {Dresselhaus}},\ }\href {\doibase 10.1142/p080} {\emph {\bibinfo {title}
  {Physical Properties of Carbon Nanotubes}}}\ (\bibinfo  {publisher} {Imperial
  College Press},\ \bibinfo {year} {1998})\BibitemShut {NoStop}%
\bibitem [{\citenamefont {Rashba}(2009)}]{rashba2009graphene}%
  \BibitemOpen
  \bibfield  {author} {\bibinfo {author} {\bibfnamefont {E.~I.}\ \bibnamefont
  {Rashba}},\ }\href@noop {} {\bibfield  {journal} {\bibinfo  {journal}
  {Physical Review B}\ }\textbf {\bibinfo {volume} {79}},\ \bibinfo {pages}
  {161409} (\bibinfo {year} {2009})}\BibitemShut {NoStop}%
\bibitem [{\citenamefont {Min}\ \emph {et~al.}(2006)\citenamefont {Min},
  \citenamefont {Hill}, \citenamefont {Sinitsyn}, \citenamefont {Sahu},
  \citenamefont {Kleinman},\ and\ \citenamefont
  {MacDonald}}]{min2006intrinsic}%
  \BibitemOpen
  \bibfield  {author} {\bibinfo {author} {\bibfnamefont {H.}~\bibnamefont
  {Min}}, \bibinfo {author} {\bibfnamefont {J.~E.}\ \bibnamefont {Hill}},
  \bibinfo {author} {\bibfnamefont {N.~A.}\ \bibnamefont {Sinitsyn}}, \bibinfo
  {author} {\bibfnamefont {B.~R.}\ \bibnamefont {Sahu}}, \bibinfo {author}
  {\bibfnamefont {L.}~\bibnamefont {Kleinman}}, \ and\ \bibinfo {author}
  {\bibfnamefont {A.~H.}\ \bibnamefont {MacDonald}},\ }\href@noop {} {\bibfield
   {journal} {\bibinfo  {journal} {Physical Review B}\ }\textbf {\bibinfo
  {volume} {74}},\ \bibinfo {pages} {165310} (\bibinfo {year}
  {2006})}\BibitemShut {NoStop}%
\bibitem [{\citenamefont {Yao}\ \emph {et~al.}(2007)\citenamefont {Yao},
  \citenamefont {Ye}, \citenamefont {Qi}, \citenamefont {Zhang},\ and\
  \citenamefont {Fang}}]{yao2007spin}%
  \BibitemOpen
  \bibfield  {author} {\bibinfo {author} {\bibfnamefont {Y.}~\bibnamefont
  {Yao}}, \bibinfo {author} {\bibfnamefont {F.}~\bibnamefont {Ye}}, \bibinfo
  {author} {\bibfnamefont {X.-L.}\ \bibnamefont {Qi}}, \bibinfo {author}
  {\bibfnamefont {S.-C.}\ \bibnamefont {Zhang}}, \ and\ \bibinfo {author}
  {\bibfnamefont {Z.}~\bibnamefont {Fang}},\ }\href@noop {} {\bibfield
  {journal} {\bibinfo  {journal} {Physical Review B}\ }\textbf {\bibinfo
  {volume} {75}},\ \bibinfo {pages} {041401} (\bibinfo {year}
  {2007})}\BibitemShut {NoStop}%
\bibitem [{\citenamefont {Gmitra}\ \emph {et~al.}(2013)\citenamefont {Gmitra},
  \citenamefont {Kochan},\ and\ \citenamefont {Fabian}}]{kochan2}%
  \BibitemOpen
  \bibfield  {author} {\bibinfo {author} {\bibfnamefont {M.}~\bibnamefont
  {Gmitra}}, \bibinfo {author} {\bibfnamefont {D.}~\bibnamefont {Kochan}}, \
  and\ \bibinfo {author} {\bibfnamefont {J.}~\bibnamefont {Fabian}},\ }\href
  {\doibase 10.1103/PhysRevLett.110.246602} {\bibfield  {journal} {\bibinfo
  {journal} {Phys. Rev. Lett.}\ }\textbf {\bibinfo {volume} {110}},\ \bibinfo
  {pages} {246602} (\bibinfo {year} {2013})}\BibitemShut {NoStop}%
\bibitem [{\citenamefont {Kochan}\ \emph {et~al.}(2017)\citenamefont {Kochan},
  \citenamefont {Irmer},\ and\ \citenamefont {Fabian}}]{kochan}%
  \BibitemOpen
  \bibfield  {author} {\bibinfo {author} {\bibfnamefont {D.}~\bibnamefont
  {Kochan}}, \bibinfo {author} {\bibfnamefont {S.}~\bibnamefont {Irmer}}, \
  and\ \bibinfo {author} {\bibfnamefont {J.}~\bibnamefont {Fabian}},\ }\href
  {\doibase 10.1103/PhysRevB.95.165415} {\bibfield  {journal} {\bibinfo
  {journal} {Phys. Rev. B}\ }\textbf {\bibinfo {volume} {95}},\ \bibinfo
  {pages} {165415} (\bibinfo {year} {2017})}\BibitemShut {NoStop}%
\bibitem [{\citenamefont {Robinson}\ \emph {et~al.}(2008)\citenamefont
  {Robinson}, \citenamefont {Schomerus}, \citenamefont {Oroszl\'any},\ and\
  \citenamefont {Fal'ko}}]{falko}%
  \BibitemOpen
  \bibfield  {author} {\bibinfo {author} {\bibfnamefont {J.~P.}\ \bibnamefont
  {Robinson}}, \bibinfo {author} {\bibfnamefont {H.}~\bibnamefont {Schomerus}},
  \bibinfo {author} {\bibfnamefont {L.}~\bibnamefont {Oroszl\'any}}, \ and\
  \bibinfo {author} {\bibfnamefont {V.~I.}\ \bibnamefont {Fal'ko}},\ }\href
  {\doibase 10.1103/PhysRevLett.101.196803} {\bibfield  {journal} {\bibinfo
  {journal} {Phys. Rev. Lett.}\ }\textbf {\bibinfo {volume} {101}},\ \bibinfo
  {pages} {196803} (\bibinfo {year} {2008})}\BibitemShut {NoStop}%
\bibitem [{\citenamefont {Liu}\ \emph {et~al.}(2011)\citenamefont {Liu},
  \citenamefont {Jiang},\ and\ \citenamefont {Yao}}]{chengLiu}%
  \BibitemOpen
  \bibfield  {author} {\bibinfo {author} {\bibfnamefont {C.-C.}\ \bibnamefont
  {Liu}}, \bibinfo {author} {\bibfnamefont {H.}~\bibnamefont {Jiang}}, \ and\
  \bibinfo {author} {\bibfnamefont {Y.}~\bibnamefont {Yao}},\ }\href {\doibase
  10.1103/PhysRevB.84.195430} {\bibfield  {journal} {\bibinfo  {journal} {Phys.
  Rev. B}\ }\textbf {\bibinfo {volume} {84}},\ \bibinfo {pages} {195430}
  (\bibinfo {year} {2011})}\BibitemShut {NoStop}%
\bibitem [{\citenamefont {Lyon}\ \emph
  {et~al.}(2017{\natexlab{a}})\citenamefont {Lyon}, \citenamefont {Sichau},
  \citenamefont {Dorn}, \citenamefont {Zurutuza}, \citenamefont {Pesquera},
  \citenamefont {Centeno},\ and\ \citenamefont {Blick}}]{lyon2017upscaling}%
  \BibitemOpen
  \bibfield  {author} {\bibinfo {author} {\bibfnamefont {T.~J.}\ \bibnamefont
  {Lyon}}, \bibinfo {author} {\bibfnamefont {J.}~\bibnamefont {Sichau}},
  \bibinfo {author} {\bibfnamefont {A.}~\bibnamefont {Dorn}}, \bibinfo {author}
  {\bibfnamefont {A.}~\bibnamefont {Zurutuza}}, \bibinfo {author}
  {\bibfnamefont {A.}~\bibnamefont {Pesquera}}, \bibinfo {author}
  {\bibfnamefont {A.}~\bibnamefont {Centeno}}, \ and\ \bibinfo {author}
  {\bibfnamefont {R.~H.}\ \bibnamefont {Blick}},\ }\href@noop {} {\bibfield
  {journal} {\bibinfo  {journal} {Applied Physics Letters}\ }\textbf {\bibinfo
  {volume} {110}},\ \bibinfo {pages} {113502} (\bibinfo {year}
  {2017}{\natexlab{a}})}\BibitemShut {NoStop}%
\bibitem [{\citenamefont {Lyon}\ \emph
  {et~al.}(2017{\natexlab{b}})\citenamefont {Lyon}, \citenamefont {Sichau},
  \citenamefont {Dorn}, \citenamefont {Centeno}, \citenamefont {Pesquera},
  \citenamefont {Zurutuza},\ and\ \citenamefont {Blick}}]{lyon2017probing}%
  \BibitemOpen
  \bibfield  {author} {\bibinfo {author} {\bibfnamefont {T.~J.}\ \bibnamefont
  {Lyon}}, \bibinfo {author} {\bibfnamefont {J.}~\bibnamefont {Sichau}},
  \bibinfo {author} {\bibfnamefont {A.}~\bibnamefont {Dorn}}, \bibinfo {author}
  {\bibfnamefont {A.}~\bibnamefont {Centeno}}, \bibinfo {author} {\bibfnamefont
  {A.}~\bibnamefont {Pesquera}}, \bibinfo {author} {\bibfnamefont
  {A.}~\bibnamefont {Zurutuza}}, \ and\ \bibinfo {author} {\bibfnamefont
  {R.~H.}\ \bibnamefont {Blick}},\ }\href@noop {} {\bibfield  {journal}
  {\bibinfo  {journal} {Physical Review Letters}\ }\textbf {\bibinfo {volume}
  {119}},\ \bibinfo {pages} {066802} (\bibinfo {year}
  {2017}{\natexlab{b}})}\BibitemShut {NoStop}%
\bibitem [{\citenamefont {Mani}\ \emph {et~al.}(2012)\citenamefont {Mani},
  \citenamefont {Hankinson}, \citenamefont {Berger},\ and\ \citenamefont
  {De~Heer}}]{mani2012observation}%
  \BibitemOpen
  \bibfield  {author} {\bibinfo {author} {\bibfnamefont {R.~G.}\ \bibnamefont
  {Mani}}, \bibinfo {author} {\bibfnamefont {J.}~\bibnamefont {Hankinson}},
  \bibinfo {author} {\bibfnamefont {C.}~\bibnamefont {Berger}}, \ and\ \bibinfo
  {author} {\bibfnamefont {W.~A.}\ \bibnamefont {De~Heer}},\ }\href@noop {}
  {\bibfield  {journal} {\bibinfo  {journal} {Nature Communications}\ }\textbf
  {\bibinfo {volume} {3}},\ \bibinfo {pages} {996} (\bibinfo {year}
  {2012})}\BibitemShut {NoStop}%
\bibitem [{\citenamefont {Wilamowski}\ \emph {et~al.}(2007)\citenamefont
  {Wilamowski}, \citenamefont {Malissa}, \citenamefont {Sch\"affler},\ and\
  \citenamefont {Jantsch}}]{wilamowski}%
  \BibitemOpen
  \bibfield  {author} {\bibinfo {author} {\bibfnamefont {Z.}~\bibnamefont
  {Wilamowski}}, \bibinfo {author} {\bibfnamefont {H.}~\bibnamefont {Malissa}},
  \bibinfo {author} {\bibfnamefont {F.}~\bibnamefont {Sch\"affler}}, \ and\
  \bibinfo {author} {\bibfnamefont {W.}~\bibnamefont {Jantsch}},\ }\href
  {\doibase 10.1103/PhysRevLett.98.187203} {\bibfield  {journal} {\bibinfo
  {journal} {Phys. Rev. Lett.}\ }\textbf {\bibinfo {volume} {98}},\ \bibinfo
  {pages} {187203} (\bibinfo {year} {2007})}\BibitemShut {NoStop}%
\bibitem [{\citenamefont {Serrano}\ \emph {et~al.}(2000)\citenamefont
  {Serrano}, \citenamefont {Cardona},\ and\ \citenamefont {Ruf}}]{cardona}%
  \BibitemOpen
  \bibfield  {author} {\bibinfo {author} {\bibfnamefont {J.}~\bibnamefont
  {Serrano}}, \bibinfo {author} {\bibfnamefont {M.}~\bibnamefont {Cardona}}, \
  and\ \bibinfo {author} {\bibfnamefont {T.}~\bibnamefont {Ruf}},\ }\href
  {\doibase https://doi.org/10.1016/S0038-1098(99)00491-3} {\bibfield
  {journal} {\bibinfo  {journal} {Solid State Communications}\ }\textbf
  {\bibinfo {volume} {113}},\ \bibinfo {pages} {411 } (\bibinfo {year}
  {2000})}\BibitemShut {NoStop}%
\bibitem [{\citenamefont {Hermann}\ and\ \citenamefont
  {Skillman}(1963)}]{hermann}%
  \BibitemOpen
  \bibfield  {author} {\bibinfo {author} {\bibfnamefont {F.}~\bibnamefont
  {Hermann}}\ and\ \bibinfo {author} {\bibfnamefont {S.}~\bibnamefont
  {Skillman}},\ }\href@noop {} {\emph {\bibinfo {title} {Atomic Structure
  Calculations}}}\ (\bibinfo  {publisher} {Prentice-Hall, Englewood Cliffs,
  NJ},\ \bibinfo {year} {1963})\BibitemShut {NoStop}%
\end{thebibliography}%
\end{document}